\documentclass[floatfix, 12pt]{revtex4}
\usepackage[dvips,dvipdfmx]{graphicx}
\usepackage{amsmath}
\usepackage{amssymb}
\usepackage{amsfonts}
\usepackage{bmpsize}

\usepackage[english]{babel}
\usepackage{hyperref}
\usepackage{color}
\usepackage[utf8]{inputenc}

\newcommand{\nc}{\newcommand}
\nc{\be}{\begin{eqnarray}}
\nc{\ee}{\end{eqnarray}}

\def\beq{\begin{equation}}
\def\eeq{\end{equation}}
\def\beqa{\begin{eqnarray}}
\def\eeqa{\end{eqnarray}}
\def\pref#1{(\ref{#1})}

\begin{document}

\title{Light dark matter around 100 GeV from the inert doublet model}
\author{Shehu AbdusSalam}
\author{Leila Kalhor}
\author{Mohammad Mohammadidoust}
\affiliation{Department of Physics, Shahid Beheshti University, Tehran, Islamic Republic of Iran}

\begin{abstract}
  We made global fits of the inert Higgs doublet model (IDM) in the light of collider and dark matter search limits and the requirement for a strongly first-order electroweak phase transition (EWPT). These show that there are still IDM parameter spaces compatible with the observational constraints considered. In particular, the data and theoretical requirements imposed favour the hypothesis for the existence of a scalar dark matter candidate around 100 GeV. This is mostly due to the pull towards lower masses by the EWPT constraint. The impact of electroweak precision measurements, the dark matter direct detection limits, and the condition for obtaining a strongly enough first-order EWPT, all have strong dependence, sometimes in opposing directions, on the mass splittings between the IDM scalars. 
\end{abstract}

\maketitle

\tableofcontents

\section{Introduction}
The observed matter-anti-matter asymmetry and dark matter (DM)
constituent of the universe are decisive indications for physics
beyond the standard model (SM) of particle physics. The theoretical
and model building developments for addressing these include the
electroweak baryogenesis scenario~\cite{Sakharov:1967dj, Kuzmin:1985mm} which
requires that earlier, the universe undergoes a strong first-order
electroweak phase transition. Particles that are stable and
weakly, or indirectly, coupled to the SM particles but with acceptable
relic densities can be considered for explaining the DM part of the 
universe~\cite{Scherrer:1985zt}.

Within the SM, there is one complex Higgs doublet which led to the
prediction of the now observed Higgs boson. Extensions of the SM Higgs
sector with additional $SU(2)_L$ n-tuples provide interesting
theoretical scenarios that could simultaneously account for
baryogenesis via a strong first-order electroweak phase transition(EWPT) in the early universe
and the observed DM density. There are many considerations with
various DM and EWPT phenomenology perspectives for addressing
non-minimal Higgs sectors. For instances, see the non-exhaustive
selection~\cite{Silveira:1985rk, Burgess:2000yq, Cirelli:2005uq,Aoki:2009vf, Cheung:2012xb, Morrissey:2012db, Blinov:2015vma, Belyaev:2016lok, Chiang:2018gsn, GAMBIT:2018eea,Chao:2018xwz, Liu:2020dok, Bandyopadhyay:2021ipw, Fan:2022dck, Khan:2015ipa, Datta:2016nfz}, the references therein and their citations.

In~\cite{AbdusSalam:2013eya}, various models extending the SM Higgs sector
using different Higgs multiplet representations were compared based on the EWPT and
DM constraints. The analyses showed that the inert Higgs doublet model(IDM)
turned out to be the favoured model. The IDM, an extension of the SM 
 by a second Higgs doublet with no direct couplings to fermions,
is one of the simplest scenarios with which a strong first-order EWPT can be
  realised and at the same time provide a candidate DM particle. 
  With a $\mathbb{Z}_{2}$ symmetry imposed, the
  lightest $\mathbb{Z}_{2}$-odd particle will be stable and hence a
  suitable DM candidate~\cite{Deshpande:1977rw} with thermal relics
  that could explain the observed DM density.
Many groups have analysed the IDM in the context of DM, EWPT, and 
collider phenomenology such as in~\cite{Fromme:2006cm, Cao:2007rm, Miao:2010rg, Honorez:2010re,Ilnicka:2015jba, Chowdhury:2011ga, Borah:2012pu, Gil:2012ya, AbdusSalam:2013eya,    Cline:2013bln, Goudelis:2013uca, Gustafsson:2012aj, Swiezewska:2013uya,   Dolle:2009ft, Blinov:2015vma, Belyaev:2016lok, Chiang:2018gsn,  Dercks:2018wch, Chao:2018xwz, Banerjee:2019luv, Fabian:2020hny,  Aoki:2021oez}.

In this  paper, we are going to make the first statistically convergent
 Bayesian global fit of the IDM in
the light of the requirement that the inert Higgs particle
simultaneously lead to a strong first-order EWPT and accounts for the
observed cold DM relic density. This should be complementary to the
work reported 
in~\cite{Eiteneuer:2017hoh}, where a frequentist global fit analysis
of the IDM were made using constraints including the DM indirect
detection limits, which we do not consider here. The requirement for a
strong first-order EWPT is central to the analysis presented here but
was not addressed in~\cite{Eiteneuer:2017hoh}.
Further, for our analysis, we have derived the IDM Lagrangian from the most general 
renormalisable potential proposed in~\cite{Chao:2018xwz} which differs from previous
general potentials used within the literature. Ultimately, we wish to derive the
Lagrangians for other representations from this and compare with the IDM to go beyond~\cite{AbdusSalam:2013eya}. 
Using {\sc LanHEP}~\cite{Semenov:2014rea}, the model 
Lagrangian was written in the form required by 
{\sc micrOMEGAs}~\cite{Belanger:2013oya,Belanger:2010pz,Belanger:2008sj, Belanger:2006is}
for computing DM properties and another form required by {\sc BSMPT} (Beyond the Standard Model 
Phase Transitions), a tool for computing beyond the SM (BSM) electroweak phase transitions~\cite{Basler:2018cwe}.

We found that the collider, DM, and theoretical constraints applied to the IDM reveal strong support for
the existence of an inert Higgs boson around 100 GeV. The most important of the constraints, namely, the 
oblique parameters from electroweak precision measurements, the dark matter direct detection limits, and the
condition for obtaining a strongly enough first-order EWPT, all have a strong dependence on the mass splittings
between the IDM scalars, $\Delta m_i$. A deeper study of the correlations with $\Delta m_i$ is an interesting
direction beyond the scope of the fits presented here which we hope to address in another project. 

The layout of this paper is as follows. In section~\ref{idm_and_constraints}, we
introduce the inert doublet model as a simple 
  extension of the standard model with one additional Higgs doublet
  $Q$ and an unbroken $\mathbb{Z}_{2}$ symmetry under which $Q$ is odd
  while all other fields are even. This discrete symmetry prevents the
  direct coupling of $Q$ to fermions and, crucial for dark matter,
  guarantees the stability of the lightest odd particle. In
  section~\ref{subsec:constraints}, we describe the theoretical conditions
  that the IDM parameters must satisfy in order to be acceptable. The
  constraints from collider searches, DM-related limits and the requirement
  for a strong first-order EWPT are also described in that section. 
  In section~\ref{sec:fits}, we present the result of the global fits
and analyses of the IDM parameter space. Our conclusions are presented in the last section.

\section{The inert doublet model (IDM)} \label{idm_and_constraints}
The $SU(2)_L \times U(1)_Y$ gauge group representations 
are labelled by isospin and hypercharge, $(J,Y)$. $J$ takes integer and
semi-integer values, and $Y$ can have any real value. The electric charge
of each component of the multiplet is given by 
$Q =T_3 + \frac{Y}{2}$. Here, $T_3$ is the third component of $SU(2)_L$
group generators that can take $n = 2J+1$ values $T_3 = J, J-1, ..., -J$ in
the {\bf n} representation. In order for one of the components to be a DM candidate, its electric
charge must be zero. This constrains the
possible values of the hypercharge, $Y$, for each $J$. 
For even(odd) values of {\bf n}, the value of $Y$ must be
an odd (even) integer and it is necessary that $|Y| \leqslant 2J$. 
For the IDM, ${\bf n} = 2$, there is only one value for hypercharge, $|Y| = 1$.
We only consider representations with a positive value of Y. Representations 
with a negative value of Y are similar to the positive ones. 

In~\cite{Chao:2018xwz}, the most general 
renormalisable scalar potential, V, with the SM Higgs
doublet, H, and an electroweak multiplet Q of arbitrary $SU(2)_L$ rank
and hypercharge, Y, was developed. 
Imposing a discrete $\mathbb{Z}_{2}$ symmetry, under which $Q$ is odd
while all the SM fields are even, prevents the lightest
$\mathbb{Z}_{2}$-odd particle from decaying into SM particles. Thus,
it could play the role of the DM candidate. 
Specialising to the IDM case, the scalar potential is given by
\begin{align} \label{eqn:Doublet}
    V = \, & \mu_h^2|H|^2 + \lambda_h|H|^4 + \mu^2_{Q}Q^{\dagger}Q + \lambda_1
    [(QQ)_1(\overline{Q}\,\overline{Q})_1]_{0}\, +
    \alpha(H^{\dagger}H)(Q^{\dagger}Q) +
    \beta[(\overline{H}H)_{1}(\overline{Q}Q)_{1}]_{0} \nonumber \\
    &+ \Big\{\kappa_1[(HH)_{1}(\overline{Q}\,\overline{Q})_{1}]_{0} +
    H.c.\Big\}
\end{align}
\begin{equation}
\textrm{ with }  H \equiv    
  \begin{pmatrix}
    H^{+} \\
    H^{0}
  \end{pmatrix}
  =\frac{1}{\sqrt{2}}
  \begin{pmatrix}
    x_1+i x_2 \\
    h + i x_3
  \end{pmatrix}
  \quad  \textrm{ and }   \quad
    Q 
  = \frac{1}{\sqrt{2}}
  \begin{pmatrix}
    y_1+i y_2 \\
    S + i R
  \end{pmatrix}.
\end{equation}
Here $\mu_{Q}, \lambda_1, \alpha, \beta$, and $\kappa_1 \equiv K$ represent the free parameters
of the IDM; $h$ is the SM-like neutral Higgs field, with the vacuum expectation value (VEV) 
$ \left< h \right> \equiv v \approx 246 \, \textrm{GeV}$; and 
$x_1$, $x_2$, and $x_3$ are the electroweak Goldstone
bosons. In unitary gauge, the parameters used here map to the commonly used $\lambda_{1,2,\dots, 5}$ notation~\cite{Belyaev:2016lok} as follows:
\begin{equation}
\left( \lambda_h, \quad \frac{1}{\sqrt{3}} \, \lambda_1, \quad \alpha, \quad  \frac{1}{2\sqrt{3}} \, \beta, \quad
\frac{2}{\sqrt{3}} \, \kappa_1\right) \rightarrow
\left( \lambda_1, \lambda_2, \lambda_3, \lambda_4, \lambda_5 \right).
\end{equation}
$\overline{H}$ and $\overline{Q}$ are the similarity transformation-related equivalents of the $SU(2)_L$
representations for $H$ and $Q$ respectively. For the IDM, with
$J =\frac{1}{2}$, the similarity transformation matrix V is equal to 
$\begin{pmatrix}
    0 & 1 \\
    -1 & 0
\end{pmatrix}$
so that
\begin{equation}
  \overline{H}= V H^{*}=\frac{1}{\sqrt{2}}
  \begin{pmatrix}
    h - i x_3\\
    -x_1+i x_2
  \end{pmatrix}
  \quad  \textrm{ and }  \quad
  \overline{Q}= V Q^{*}=\frac{1}{\sqrt{2}}
  \begin{pmatrix}
    S - i R \\
    -y_1+i y_2
  \end{pmatrix}.
\end{equation}\\
This way, $(\overline{Q}_{j=\frac{1}{2}}\,Q_{j=\frac{1}{2}})_J$
represents the combination of two $j=\frac{1}{2}$ doublets with total isospin $J$. The isospin addition rules should be used.
For instance
\begin{equation}
(\overline H H)_1 = 
 \begin{pmatrix}
    (H^0)^* H^+ \\
    \frac{1}{\sqrt{2}}\left[( H^0)^*H^0 - (H^+)^* H^+ \right] \\
    -(H^+)^* H^0
  \end{pmatrix}. 
\end{equation}

The mass terms for the neutral scalar $S$, the pseudoscalar state, $R$, and for the charged scalar, $Y^{\pm}$, after electroweak symmetry breaking are
\begin{equation}\label{eqn:MDM}
  M_{S}^2 = \mu_Q^2 +(\frac{\alpha}{2}+\frac{\sqrt 3}{3}K -
  \frac{\sqrt 3}{12}\beta)v^2,  \, 
M_{R}^2 = \mu_Q^2 +(\frac{\alpha}{2}-\frac{\sqrt 3}{3}K - \frac{\sqrt
  3}{12}\beta)v^2,  \, 
M_{Y^{\pm}}^2 = \mu_Q^2 +(\frac{ \alpha}{2}+\frac{\sqrt
  3}{12}\beta)v^2. 
\end{equation}
For $S$ to be the DM candidate particle, and stable, $M_S < M_R , M_{Y^{\pm}}$ must be satisfied. Accordingly, this
choice will imply that $K<0$ and $\frac{\sqrt 3}{3}K+\frac{\sqrt  3}{6}\beta<0$. Using the Higgs portal notations, 
\begin{equation}
\Lambda_1 = \frac{\alpha}{2}+\frac{\sqrt 3}{3}K - \frac{\sqrt 3}{12}\beta,
  \quad \textrm{ and } \quad \bar\Lambda_1 = \frac{\alpha}{2}-\frac{\sqrt 3}{3}K - \frac{\sqrt 3}{12}\beta
\end{equation}
are respectively related to the triple and quartic couplings between the SM Higgs $h$ and the DM candidate $S$ or the pseudoscalar $R$.
The parameters $\alpha$ and $\beta$, on the other hand, determines the mass term, and describe the $h$ interaction with the charged
scalars $Y^{\pm}$. The parameter $\lambda_1$ describes the quartic self- and non-self couplings of extended Higgs sector particles.
The vertex factors are summarised in Table~\ref{tab:table1}. 
The portal parameters can be expressed in terms of the mass parameters and $\lambda_1$ as follows
\begin{equation}
 \alpha = 2 \Lambda_1 - \frac{1}{2v^2}(3M_S^2-2M_{Y^\pm}^2-M_R^2), \, \beta 
 = -\frac{\sqrt 3}{v^2}(M_S^2+M_R^2-2M_{Y^\pm}^2), \, K = \frac{\sqrt 3}{2v^2}(M_S^2-M_R^2). 
\end{equation}
\begin{table}
  \begin{center}
    \caption{The vertex factors for the IDM couplings.}
    \label{tab:table1}
    \begin{tabular}{|l|l||l|l|}
      \hline
      \textbf{vertex} & \textbf{factor} & \textbf{vertex} & \textbf{factor} \\
      \hline
      $SSSS$ & $2\sqrt3\lambda_1$ & $hhY^+Y^-$ & $\alpha+\frac{\sqrt3}{6}\beta$ \\
      $RRRR$ & $2\sqrt3\lambda_1$ & $hY^+Y^-$ & $(\alpha+\frac{\sqrt3}{6}\beta)v$ \\
      $Y^+Y^-Y^+Y^-$ & $4\frac{\sqrt3}{3}\lambda_1$ & $hhRR$ & $2\bar\Lambda_1$ \\
      $SSRR$ & $\frac{2\sqrt3}{3}\lambda_1$       & $hSS$ & $2\Lambda_1 v$ \\
      $SSY^+Y^-$ & $\frac{2\sqrt3}{3}\lambda_1$ & $hhSS$ & $2\Lambda_1$ \\
      $RRY^+Y^-$ & $\frac{2\sqrt3}{3}\lambda_1$ & {} &\\
      \hline
    \end{tabular}
  \end{center}
\end{table}

In all, the IDM has five free parameters which can be chosen to be $M_S, M_R, M_{Y^{\pm}}, \lambda_1$, and $\Lambda_1$.
For the global fits, the mass parameters were allowed in the range, 1 to 5 TeV. The parameter $\Lambda_1$ was allowed in [-1, 1] while 
$\lambda_1$ was fixed at 0.1. In what follows, we are going to explain the set of theoretical and experimental results used for 
constraining the IDM parameters space. 

\section{The constraints on the IDM} \label{subsec:constraints} 
\subsection{Theoretical constraints}

\paragraph{Vacuum Stability:}
A scalar potential has to be bounded from below for describing a stable physical 
system. Within the SM, this 
means that the self-coupling of the Higgs boson, $\lambda_h$, has to be
positive. For the IDM, the vacuum of the potential has to be stable  in the
limit of large  values along  all possible
directions of the field space. This will require that~\cite{Kannike:2012pe}
\begin{equation}
  \lambda_h > 0, \, \lambda_1 > 0; \, \alpha + 2\sqrt{\lambda_1
    \lambda_h}>0; \, \textrm{ and } 
  \alpha+\frac{\sqrt 3}{6}(\beta-|K|)+2\sqrt{\lambda_1
    \lambda_h}>0.  
\end{equation}  

\paragraph{Perturbativity and unitarity:}
{For calculations using perturbation theory, the relevant couplings
used as expansion parameters should not be too large. This can be
imposed by requiring that the absolute values of the coupling
parameters be less than $4\pi$~\cite{Ginzburg:2005dt}.}
We also require that unitarity should not be violated for all
  scalar $2 \rightarrow 2$ scattering. The perturbative unitarity
  conditions~\cite{Ginzburg:2005dt, Belyaev:2016lok} applied to the
  IDM are $|U_i| \leq 8 \pi$, where $i = 1, \ldots 10$, 
\begin{eqnarray}
  U1 &=& \lambda_h + \frac{\sqrt 3}{3}\lambda_1 + \frac{\sqrt
    {3\lambda_1^2 - 6{\sqrt 3}\lambda_1 \lambda_h+3 \beta^2+9
      \lambda_h^2}}{3},
  \quad U5 = \alpha - \frac{{\sqrt 3}}{6}\beta -
  4 \frac{\sqrt 3}{3}K \nonumber\\
  U2 &=& \lambda_h + \frac{\sqrt 3}{3}\lambda_1 - \frac{\sqrt
    {3\lambda_1^2 - 6{\sqrt 3}\lambda_1 \lambda_h+3 \beta^2+9
      \lambda_h^2}}{3},
  \quad U6 = -\alpha + \frac{{\sqrt 3}}{6}\beta + 4 \frac{\sqrt
    3}{3}K, \nonumber\\ 
  U3 &=& \lambda_h + \frac{\sqrt 3}{3}\lambda_1 + \frac{\sqrt
  {3\lambda_1^2 - 6{\sqrt 3}\lambda_1 \lambda_h+3 K^2+9
    \lambda_h^2}}{3},
  \quad U7 = \alpha - \frac{{\sqrt 3}}{6}\beta, \nonumber\\ 
  U4 &=& \lambda_h + \frac{\sqrt 3}{3}\lambda_1 - \frac{\sqrt
    {3\lambda_1^2 - 6{\sqrt 3}\lambda_1 \lambda_h+3 K^2+9
      \lambda_h^2}}{3},
  \quad U8 =  \alpha + \frac{\sqrt 3}{2} \beta,\nonumber\\
  U9 &=& \alpha + \frac{\sqrt 3}{6} \beta + \frac{\sqrt 6}{3}K,
  \quad \textrm{ and }
  \quad U10 = \alpha + \frac{\sqrt 3}{6} \beta - \frac{\sqrt 6}{3} K.  
\end{eqnarray}

\subsection{Limits from collider searches}
The main approach for the phenomenological exploration of BSMs is the confrontation of the models with limits from
  experiments. The large
electron-positron(LEP), Tevatron and the large hadron collider (LHC) experiments publish exclusion
  limits based on precision measurements or the non-observation of new
  particles. For exploring and fitting the IDM parameter space to data, the categories of collider limits used are explained as
  follows. 
\paragraph{LEP:} 
Precision measurement results by LEP exclude the possibility that
massive SM gauge bosons decay into inert particles. This requires
that~\cite{Dercks:2018wch, Cao:2007rm, Belyaev:2016lok} 
\begin{equation}
M_{R,S} + M_{Y^\pm} \geq M_W, \quad M_R + M_S \geq M_Z, \textrm{ and } 2M_{Y^\pm} \geq
M_Z.
\end{equation}
{The LEP results also} give rise
to the exclusion of an intersection of mass ranges which can
be evaded by fulfilling all of the following conditions
{simultaneously}~\cite{Belyaev:2016lok,Ilnicka:2015jba} 
\begin{equation}
\label{eq:LEPsplit}
M_{S} > 80 GeV, \quad M_{R} > 100 GeV, \quad M_{R}-M_{S} < 8 GeV. 
\end{equation}
{There is also the limit $M_{Y^{\pm}} > 70 GeV$ } 
from searches for charged Higgs pair production.

\paragraph{Oblique parameters:}
The values of the S, and T (with U=0) oblique parameters for a given point in the IDM parameter space can be computed
as~\cite{hep-ph/0603188, Belyaev:2016lok}
\begin{equation}
\label{STUeq}
\begin{split}
S & =  \frac{1}{72\pi \left( x_{2}^{2} - x_{1}^{2} \right)^{3}} \left[ x_{2}^{6}f_{a}\left( x_{2} \right) - x_{1}^{6} f_{a} \left( x_{1} \right) + 9x_{1}^{2}x_{2}^{2} \left( x_{2}^{2}f_{b} \left( x_{2} \right) - x_{1}^{2}f_{b} \left( x_{1} \right) \right) \right]\\
T & =  \frac{1}{32\pi^{2}\alpha v^{2}} \left[ f_{c} \left( M_{Y^{\pm}}^{2}, M_{R}^{2} \right) + f_{c} \left( M_{Y^{\pm}}^{2}, M_{S}^{2} \right) - f_{c} \left( M_{R}^{2}, M_{S}^{2} \right)  \right] \\
&\simeq   \frac{1}{24\pi^{2}\alpha v^{2}}(M_{Y^\pm}-M_S)(M_{Y^\pm}-M_R)\,
= \frac{\Delta m_{Y^\pm} \, ( \Delta m_{Y^\pm} - \Delta m_R)}{24\pi^{2}\alpha v^2}\,.
\end{split}
\end{equation}
Here, $\Delta m_{Y^\pm} = M_{Y^\pm}-M_S$,  $\Delta m_R = M_R-M_S$, $x_{1} \equiv \frac{M_{S}}{M_{Y^{\pm}}} $,  $ x_{2} \equiv \frac{M_{R}}{M_{Y^{\pm}}} $, $\alpha\approx 1/127$ denotes the fine-structure constant at the scale of the $Z$ boson mass, $f_{a}(x) \equiv -5 + 12\ln x$, $f_{b}(x) \equiv 3-4\ln x$, and 
\begin{equation}
\label{eq:fs}
f_{c}(x,y) \equiv \begin{cases} \frac{x+y}{2} - \frac{xy}{x-y}\ln \frac{x}{y} &\mathrm{for} \ x\neq y \\ 0 &\mathrm{for} \ x=y \end{cases}\,.
\end{equation}
Only model points with S and T oblique parameters within 1-sigma of the PDG~\cite{ParticleDataGroup:2020ssz} average were accepted. There
is a strong correlation between these parameters and the limit on other observables used for the IDM analyses. In particular, both the
requirement for a strong first-order EWPT $\frac{v_c}{T_c} > 1$ and the upper bound on the inert singlet versus nucleons elastic
scattering cross section, $\sigma^{SI}$, are strongly dependent on the IDM mass differences. This is also the case for the oblique parameters
as can be seen in Eq.(\ref{STUeq}). The new precision measurements of the top-quark and W boson masses~\cite{CDF:2022hxs, CMS:2022kcl} will
change the allowed ranges for the oblique parameters. We check that the IDM fit presented here are compatible with these within 2-sigma
range of the allowed interval from an updated global fit~\cite{2204.04204}, which includes these new measurements, of new physics
models to electroweak precision data. 

\paragraph{Limits implemented in HiggsBounds:} The Higgs sector
predictions based on the IDM are compared with corresponding cross
section limits for various processes studied at LEP, Tevatron, and LHC to determine whether the IDM parameter point has been
excluded at 95\% C.L. or not. HiggsBounds~\cite{0811.4169, 1102.1898, 2006.06007} incorporates results from
LEP~\cite{hep-ex/0107034, hep-ex/0107032,  hep-ex/0107031,hep-ex/0111010, CERN-ALEPH-2002-019,  hep-ex/0206022, hep-ex/0401022,  hep-ex/0401026, hep-ex/0404012, hep-ex/0501033, hep-ex/0410017,  hep-ex/0602042, 0707.0373, 0812.0267, 1301.6065}, the Tevatron~\cite{0809.3930, 0806.0611, 0908.1811, 0907.1269, 0906.1014,   0905.3381, 1011.1931, 1001.4468, 1001.4481, 1003.3363, 1008.3564,  1107.1268, 1106.4555, 1106.4885, 1108.3331, 1203.3774, 1207.6436}, the ATLAS~\cite{1207.7214, 1112.2577, 1109.3357, 1108.5064,   1202.1415, 1202.1414, 1202.1408,  1204.2760, 1402.3051, 1402.3244, 1407.6583, 1409.6064, 1406.7663,  1406.5053, 1509.00389, 1509.05051, 1507.05930, 1503.04233, 1502.04478, 1509.04670,  1606.04833,  1606.08391,  1710.07235,  1710.01123, 1712.06386, 1709.07242, 1707.04147, 1808.02380,  1808.03599, 1804.01126, 1807.00539, 1807.08567, 1807.07915,  1806.07355, 1811.11028, 1809.06682, 1808.00336, 1909.10235, 1904.05105, 1901.08144, 1907.06131, 1906.02025,   1907.02749}, and the CMS~\cite{1202.1997,   1202.3478, 1202.1416, 1202.1488, 1312.5353, 1307.5515,  1310.3687, 1407.0558, 1404.1344, 1504.00936,  1504.04710, 1510.06534, 1506.02301,  1508.07774, 1510.01181, 1506.08329, 1510.04252, 1506.00424,  1503.04114, 1603.02991, 1603.06896, 1707.02909, 1708.04188,  1701.02032, 1707.07283, 1811.08459, 1808.06575,  1805.04865, 1812.06359, 1809.05937, 1805.10191, 1804.01939,  1805.12191, 1811.09689, 1803.06553, 1911.04968, 1907.07235,  1903.04560, 1912.01594, 1911.10267, 1907.03152,   1908.01115, 1911.03781, 1903.00941, 2001.07763}
experiments. 

\paragraph{Limits implemented in {\sc Lilith}:}
Should the second CP-even Higgs boson of the IDM be SM-like, with mass
between 123 to 128 GeV, then {\sc Lilith}~\cite{1502.04138,1606.03834,1908.03952}
is used for gauging its couplings with respect to the Higgs signal
strength measurements from
ATLAS~\cite{Aaltonen:2013xpo, Aad:2015gba, Aad:2014eha,    Aad:2015ona, Aad:2014eva, Aad:2015vsa, Aad:2015iha, Aad:2015gra,    Aad:2014xva,  Aad:2014iia,    Aaboud:2018xdt, Aaboud:2017vzb, Aaboud:2018pen, Aaboud:2017ojs, Aaboud:2018jqu, Aaboud:2018gay, Aaboud:2019rtt, Aad:2019lpq, Aaboud:2017xsd, Aaboud:2017bja, Aaboud:2017jvq, Aaboud:2017rss} and  CMS~\cite{Khachatryan:2014jba, Khachatryan:2014ira,     Chatrchyan:2013iaa, Chatrchyan:2013mxa, Chatrchyan:2014nva,    Chatrchyan:2013zna, Khachatryan:2014qaa, Khachatryan:2015ila,    Khachatryan:2015bnx, Chatrchyan:2014tja, Sirunyan:2018koj,Sirunyan:2018cpi, Sirunyan:2018owy}.
 For each IDM point with an associated signal strength $\mu_i$ , {\sc
   Lilith} returns a log-likelihood value 
\be
-2L_{lilith}( \theta) = - 2 \sum_i \log L(\mu_i) = \sum_{i}
\left(\frac{\mu_i(\theta) - \hat\mu_i}{\Delta \mu_i}\right)^2. 
\ee
Here $i$ runs over the various categories of Higgs boson production
and decay modes combinations for a given point, $\theta$, in the model
parameter space. $\hat\mu_i \pm \Delta\hat\mu_i$
represents the experimentally determined signal strengths. 
Theoretically, the signal strength associated to a model point for a
given production mode $X$ and decay mode $Y$ is 
\begin{equation} \label{muimp}
\mu = \sum_{X,Y} \epsilon_{X,Y} \frac{ \sigma(X) \, BR(H \rightarrow Y)}{\left[ \sigma(X) \, BR(H \rightarrow Y) \right]^{SM}}
\end{equation}
where $\epsilon_{X,Y}$ represents experimental efficiencies,
$X \in \{ ggH, VH, VBF, ttH\}$ and
$Y \in \{ \gamma \gamma, VV^{(*)}, b \bar{b}, \tau \tau, ttH\}$. 
In general, for the results from LHC, the elements in $X$ represent: 
the gluon-gluon fusion (ggH), associated production with a boson (VH), 
vector boson fusion (VBF) or associated production with top quarks 
(ttH). The elements in $Y$ represent the Higgs diphoton
($\gamma \gamma$), W or Z bosons ($VV$), bottom quarks ($bb$) or tau
leptons ($\tau \tau$) decay modes.

For computing the signal strengths $\mu$, the input parameters passed
to {\sc Lilith} are the reduced couplings~\cite{Heinemeyer:2013tqa}
$C_X^2$ and $C_Y^2$ such that  
\begin{equation} 
\sigma(X) =  C_X^2 \, \sigma(X)^{SM} \quad \textrm{ and } \quad 
\Gamma(Y) =  C_Y^2 \, \Gamma(Y)^{SM}.  
\end{equation}
These, together with the Higgs boson invisible and undetectable decay
branching ratios are computed using the {\sc micrOMEGAs} system for 
\begin{equation}
\mu = ( 1 - BR(H \rightarrow  \, undetected) - BR(H \rightarrow \,
invisible) ) \frac{ \sum_{X,Y} \epsilon_{X,Y} C_X^2 \, C_Y^2} {\sum_Y
  \, C_Y^2 \, BR(H \rightarrow Y)^{SM}}. 
\end{equation}
This is then in turn compared with the table of likelihood values as a
function of $\mu$ within the {\sc Lilith} database of results from
experiments for computing the log-likelihood. 

\subsection{Dark matter related constraints}
The IDM predicts the existence of
a neutral scalar field, S as  DM candidate. The SM Higgs boson
may decay into a pair of the DM candidate particles when
  kinematically allowed, and can therefore 
contribute to the invisible SM Higgs boson. For the IDM, we
  require that the branching ratio of the SM Higgs boson decay to the
  DM candidate particle be less than $0.15$~\cite{ATLAS-CONF-2020-052}. 

  At early universe times, after freezing-out of equilibrium, the
  relic density of $S$ can account for the observed density of DM relics.
 The scattering of S onto nucleons  should possibly lead to DM direct detection 
signatures. 
There are searches for the elastic scattering of DM with
nucleons. It is expected that the recoil energy deposited on nuclei in
a detector can be measured. In the absence of discovery, then 
upper limits on the scattering cross section can be determined. The
cross sections can be either spin-independent (SI) or spin-dependent
(SD) depending on whether the lightest odd particle effective coupling to the nucleons  
is via scalar or axial-vector interaction. The currently most stringent direct detection limits are those by 
PandaX-II~\cite{PandaX-II:2017hlx} and the XENON1T~\cite{XENON:2019gfn} experiments. 
 We use the package {\sc micrOMEGAs} for computing the IDM
  predictions for the DM candidate relic density and its scattering
  cross section while interacting with nucleons.  These are then
  compared with the corresponding experimentally determined value,
  $\Omega_{DM}h^2 = 0.1200 \pm 0.0001$~\cite{1807.06209} for the relic
  density, and direct detection limits set by
  PandaX-II~\cite{PandaX-II:2017hlx} and the XENON1T~\cite{XENON:2019gfn} experiments.

\subsection{Requirement for strong first-order EWPT}
For investigating the EWPT, the finite temperature quantum field 
  theory techniques has to be used -- see~\cite{Quiros:1999jp} for a review.
The ground state of the potential at $v= 0$ represents 
 the symmetric phase of the model, while $v \ne 0$ represents the
broken phase. Starting with the symmetric vacuum in the early
universe, the EWPT is defined as the point in the evolution of the
effective potential, $V_{eff}$, where a second minimum with non-zero
VEV, $v_c$,  develops at the critical temperature $T_c$ such that 
\begin{equation} \label{veff}
V_{eff} (v=0, T_c) = V_{eff} (v=v_c, T_c), \quad V_{eff} =
V_{tree}+V_{CW}+V_{CT}+V_T.  
\end{equation}
Here $V_{tree}$, $V_{CW}$, $V_{CT}$ and $V_T$ respectively represents the tree-level
potential, Eq.(\ref{eqn:Doublet}), the Coleman-Weinberg potential, the
counter-term potential  and the thermal corrections at finite temperature T. The
latter set of the effective potential terms are described in
Appendix~\ref{veffective}.

Given the IDM tree-level potential, the other terms above were computed 
such that the strength of the EWPT at each model point can be
determined using BSMPT~\cite{1803.02846, 2007.01725}. The description
of the IDM implementation into the BSMPT is given in
Appendix~\ref{makeBSMPT}.  
BSMPT can find the
global minimum of $V_{eff}$ and hence determine $T_c$ and
$v_c$ at the instance when the phase transition takes place. 
For the model point  to be a possible candidate for electroweak 
baryogenesis, the EWPT must be strongly first-order in order to
  suppress sphaleron wash-out within the broken phase region;  
see~\cite{Morrissey:2012db} for a review. The required condition for a strong 
first-order EWPT is~\cite{Kuzmin:1985mm}
\begin{equation}
\xi_c \equiv \frac{v_c}{T_c} > 1.
\end{equation}

\section{Results of the IDM global fits} \label{sec:fits}
\begin{figure}
  \includegraphics[width=1.\textwidth]{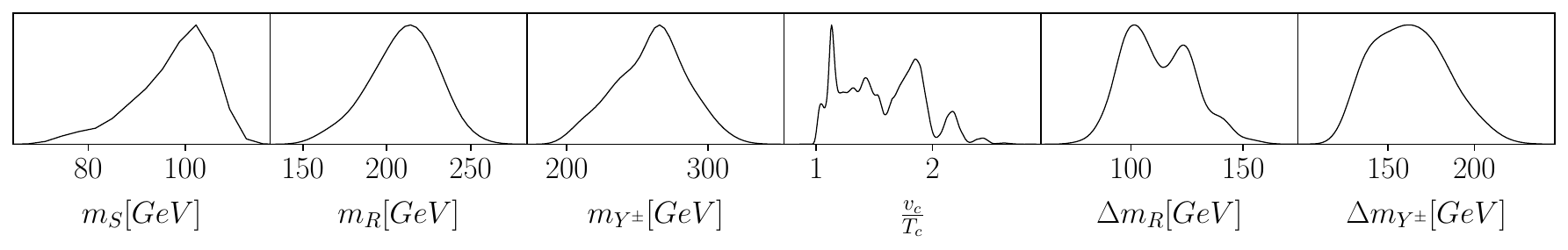} 
  \includegraphics[width=1.\textwidth]{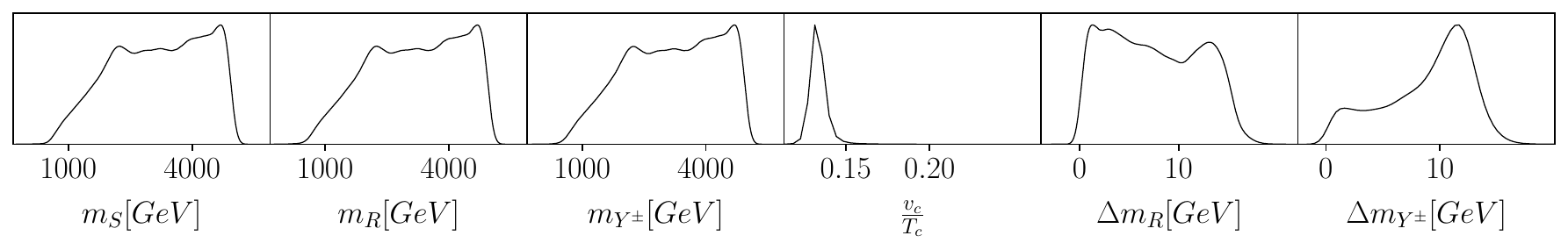}  
  \caption{1-Dimensional posterior distributions from the global fit of the IDM parameters to data. The prior distributions for the mass
    parameters are flat in the range, [1, 5000] GeV. All model points pass the set of constraints explained in
    section~\ref{subsec:constraints}. In particular the strength of the first-order EWPT, $\frac{v_c}{T_c} > 1$. The last two plots
    present the posteriors of the mass splittings between the inert scalars, $\Delta m_R = m_R - m_S$ and
    $\Delta m_{Y^{\pm}} = m_{Y^{\pm}} - m_S$. The second row of plots is the same as the first, but without $\frac{v_c}{T_c} > 1$
    imposed.}
  \label{1dposteriors}
\end{figure}

The sampling and fit of the IDM parameter space, $\theta$, with the SM
Higgs boson mass fixed, within an inflated range for accommodating theoretical uncertainties, 
at $m_h = 125 \pm 3 \, \textrm{GeV}$, 
is done using {\sc MultiNest}~\cite{Feroz:2007kg,Feroz:2008xx}. 
Only model points that pass the set of theoretical
and experimental constraints, $d$, described in
section~\ref{subsec:constraints} and for which the lightest odd
particle is the CP-even IDM Higgs boson, $S$, are passed for
implementing the nested sampling algorithm. For these IDM parameter
points, we model the likelihood $p( d| \theta)$ of the IDM
predictions, $O_i$, corresponding to the i$^{th}$ constrain, with
experimental central values $\mu_i$ and uncertainties $\sigma_i$, as   
\begin{equation} \label{likel} p( d| \theta) = \prod_i \, \frac{ \exp\left[- (O_i - \mu_i)^2/2 \sigma_i^2\right]}{\sqrt{2\pi \sigma_i^2}}.
\end{equation}   
For scenarios where $S$ is pseudo-degenerate with the SM Higgs, additional contributions based on the Higgs signal strength
measurements at colliders as implemented in {\sc Lilith} were added to~Eq.(\ref{likel}). The global fit indicates that the lightest
inert Higgs should be expected around $m_S = 97.83 \pm 11.49 \, \textrm{GeV}$. At Maximum a Posteriori (MAP) and maximal likelihood, 
$m_S \sim 100 \, \textrm{GeV}$. This result supports the possibility for the IDM inert Higgs boson account for the observed mild but
independent excesses at LEP and CMS experiments~\cite{hep-ex/0306033, CMS-PAS-HIG-14-037, CMS-PAS-HIG-17-013, 1811.08459, CMS-PAS-HIG-21-001} in search for light Higgs bosons. In Figure~\ref{1dposteriors}, the 1-dimensional posterior distributions of the IDM parameters are shown. 

\begin{figure}
  \includegraphics[width = 3.2in]{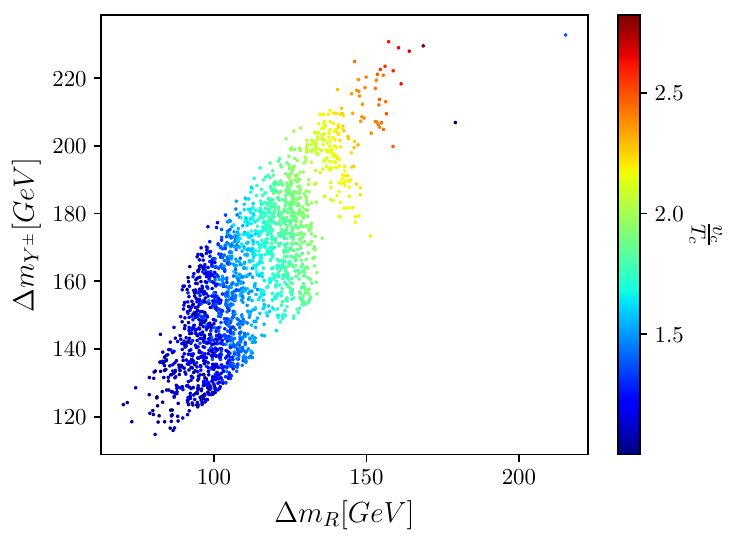}
  \includegraphics[width = 3.2in]{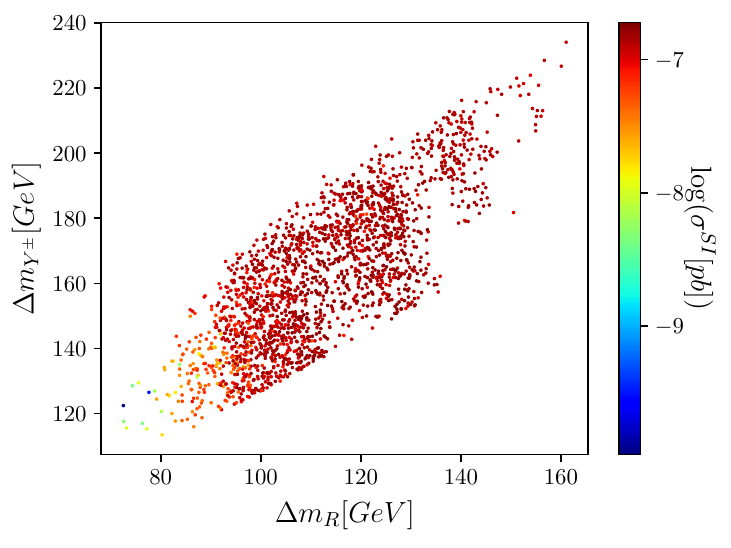}    
  \caption{Scatter plots showing the correlations between the inert scalars' mass differences with: (left) strength of the first-order
    EWPT, $\frac{v_c}{T_c}$, and (right) the inert single DM candidate particle's elastic scattering cross section against nucleons,
    $\sigma^{SI}$. Both quantities show dependence on the mass differences but in opposite directions with respect to the required
    respective limits.} 
  \label{posterior_2d_2}
\end{figure}

As is the case for the SM, strong first-order EWPT condition, $\frac{v_c}{T_c} > 1$, translates into an upper bound on the lightest inert Higgs mass.
This partly explains way a significant part of the prior region, with $m_S \sim 1 \, \textrm{to}\, 5 \, \textrm{TeV}$  will be disfavoured.
Should 
this requirement be uplifted, multi-TeV $m_S$ are possible as can be seen on the second row of Figure~\ref{1dposteriors} plots. 
Imposing the upper limit~\cite{PandaX-II:2017hlx, XENON:2019gfn} on the inert singlet versus nucleons elastic scattering cross section,
$\sigma^{SI}$, on the IDM fit to data leads to small mass differences $\Delta m_{Y^{\pm}} \sim \Delta m_R \sim {\cal O}(10) \, \textrm{GeV}$.
Contrary to this, the strength of the first-order EWPT, $\frac{v_c}{T_c}$, is proportional to the mass splittings 
$\Delta m_{R, \, Y^{\pm}}$. The tendencies with respect to the mass differences can be seen in Figure~\ref{posterior_2d_2}. Therefore, simultaneously requiring both $\frac{v_c}{T_c} > 1$ and the DM direct detection limits on the IDM 
fit to data is extremely difficult beyond the scope and the computational resources at our disposal. As such, the direct detection
limits were not imposed for final fit of the IDM to data since this extremely slowed the sampling of the model parameter space. Instead,
the impact of this limit were assessed via post-processing the posterior samples~\footnote{Including the DM direct detection limit for the
IDM fits is an interesting direction we hope to pursue in the future using machine-learning techniques.}. Complementary to post-processing,
a dedicated fit 
of the model with DM direct limit imposed but without the strong first-order EWPT requirement could be used for assessing the tension between
both observables. The second row of plots Figure~\ref{1dposteriors} were from such an 
IDM fit with the direct detection 90\% C.L. exclusions limits~\cite{XENON:2018voc, DarkSide:2018bpj,PICO:2019vsc,CRESST:2019jnq} imposed. As 
can be seen, this yields posteriors with $\frac{v_c}{T_c} <0.1$. 

The constraints from collider limits disfavours small values of $m_S$ and $m_{Y^{\pm}}$, and also contribute to the control that leads 
to the allowed region for $\Delta m_{R, \, Y^{\pm}}$. This particularly important for the oblique parameters constraints which favour 
relatively lower inert Higgs mass differences. In all, 
 there are IDM points that 
satisfy collider and dark matter searches limits including relic density generation, and simultaneously allow for a strong first-order
EWPT. A selection of benchmarks are presented in Table~\ref{benchmarks}.  
\begin{table}[h]
\centering
\begin{tabular}{|l|l|l|l|l|l|l|l|l|}
  \hline
  $m_{S}$ & $m_{R}$ & $m_{Y^{\pm}}$ & $\Lambda_1$ & $\log \, (\Omega_{DM} \, h^2)$ & $\log (\sigma^{SI} [pb])$ & $\frac{v_c}{T_c}$ & $\Delta m_{R}$  & $\Delta m_{Y^{\pm}}$  \\
 \hline
 93.6989 & 176.7693 & 213.3181 & 0.2853 & -2.10594 & -8.6287 &  1.0761 & 83.0704 & 119.619 \\
 115.0309 & 133.2492 & 209.0929 & 0.3971 & -3.64282 & -9.28878 & 1.2038 & 18.2183 & 94.0621 \\
 101.5063 & 157.1630 & 216.7150 & 0.3678 & -3.36139 & -9.72497 & 1.2374 & 55.6567 & 115.209 \\
 122.3877 & 198.1699 & 260.3158 & 0.4478 & -3.26165 & -10.8541 & 1.5127 & 75.7821 & 137.928 \\
 80.5285 & 151.8615 & 193.4262 & 0.28262 & -2.19879 & -12.0729 & 1.0923 & 71.333 & 112.898 \\
 \hline
\end{tabular}
\caption{A selection of benchmark points for the IDM model which passed all the constraints considered. Masses are in GeV units.}
\label{benchmarks}
\end{table}

\section{Conclusion} 
We have made a detailed exploration of the parameters and the interplay 
amongst them for the inert Higgs doublet model in light 
of the limits from collider experiments, the constraints from dark matter 
searches and the requirement for a strong first-order electroweak phase 
transition (EWPT). The software packages BSMPT and micrOMEGAs were used for 
computing the strength of EWPT and dark matter properties respectively. 
The collider constraints on the IDM parameter space were applied by using 
HiggsBounds and Lilith packages. Our analyses include the first global fit 
of the IDM to data using a statistically convergent Bayesian approach 
implemented in MultiNest package. For these, we have used, also for the 
first time, a recently presented most general renormalisable potential for 
the IDM. This will lead to different phase transition dynamics and coupling 
constants. 

The global fits show that the IDM spectra can have DM candidates with 
all constraints satisfied and at the same time able to produce a 
strongly first-order EWPT. A selection of benchmark points was provided which can be analysed further 
with respect to ongoing or future collider experiments. The posterior
sample~\footnote{The posterior sample can be downloaded at \url{https://doi.org/10.7910/DVN/TCMXDS}.}
could be useful 
for addressing IDM collider and dark matter phenomenology. Given
computing resources, the chain of particle physics phenomenology tools
developed can be used for comparisons amongst the extended Higgs sector BSMs
in the light EWPT and dark matter candidate particle constraints.

There are several interesting research directions that could be built on the result presented here. Besides the S and T parameters
constrained already applied for the fits, there already are interesting collider results from the LHC which
could possibly probe further the IDM parameters space. For instance, at the LHC, the electroweak production of $Y^\pm$ and $S$ (or $R$) can
subsequently lead to the inert Higgs decays into a weak boson and the lightest inert scalar. These can provide the same final states and
therefore can be constrained by limits from supersymmetric electroweakino searches~\cite{ATLAS:2021moa, ATLAS:2021yqv}. 
 Implementing such limits
on the IDM will, however, require dedicated reinterpretation studies. The fits revealed strong correlations between three important
constraints ($\sigma^{SI} [pb]$, $\frac{v_c}{T_c}$, and the electroweak precision oblique parameters S and T) used with a strong dependence
on the IDM Higgs mass differences albeit with possible pulls along opposing directions. A better delineation of the IDM with respect to
these constraints can be achieved by using machine-learning techniques in exploring the compatible regions in parameter space. 

\paragraph*{Acknowledgements:}
We would like to thank Philip Basler for his kind help with the BSMPT package, Alexander Belyaev for advice on the
LanHep package and all developers of micrOMEGAs package for their help. MM would like to thank Najimuddin Khan for discussions
about perturbative unitarity, Per Osland for the discussions about oblique parameters. This work was partly performed using resources provided by the Cambridge Service for Data Driven Discovery (CSD3) operated by the University of Cambridge Research Computing Service (www.csd3.cam.ac.uk), provided by Dell EMC and Intel using Tier-2 funding from the Engineering and Physical Sciences Research Council (capital grant EP/T022159/1), and DiRAC funding from the Science and Technology Facilities Council (www.dirac.ac.uk). 

\begin{appendix}
\section{The IDM finite temperature effective potential} 
\label{veffective}
The 1-loop finite temperature effective potential for the IDM can br
written as follows, following the BSMPT~\cite{1803.02846, 2007.01725}
notations, in terms static field configuration $\omega$ and
temperature T. 
\begin{equation}
V (\omega, T ) = V(\omega) + V_T(\omega, T ) = V_{tree}(\omega) + V_{CW}(\omega) + V_{CT}(\omega) + V_T(\omega, T )
\end{equation}  
where $V(\omega)$ consists of the tree-level potential $V_{tree}$,
Eq.\pref{eqn:Doublet}, the Coleman-Weinberg potential $V_{CW}$ and the
counter-term potential $V_{CT}$. The thermal corrections to the
potential is given by $V_T(\omega, T)$. In this section we briefly
describe each of these terms and then how they are implemented into
the BSMPT package in Appendix~\ref{makeBSMPT}.

The notation
of~\cite{Camargo-Molina:2016moz} is used for casting the effective
potential into the form
\begin{align}
-\mathcal{L}_S &= L^i \Phi_i + \frac{1}{2!} L^{ij} \Phi_i \Phi_j+\frac{1}{3!} L^{ijk} \Phi_i
\Phi_j\Phi_k + \frac{1}{4!} L^{ijkl} \Phi_i \Phi_j \Phi_k \Phi_l \label{Eq:LS}\\
-\mathcal{L}_F &= \frac{1}{2} Y^{IJk} \Psi_I \Psi_J \Phi_k + c.c. \label{Eq:LF}\\
\mathcal{L}_{G} &= \frac{1}{4} G^{abij}
A_{a\mu}A_b^\mu\Phi_i\Phi_j \;,  \label{Eq:LG} 
\end{align}    
with summation over repeated indices implied if one is up and the
other is down. In this manner, the IDM scalar multiplets are 
decomposed into $n_{\text{Higgs}}$ real scalar fields $\Phi_i$, with
$i= 1,\dots , n_{\text{Higgs}} = 8$. Here $\Psi_I$, with $I=1,\dots ,
n_{\text{fermion}}$ represents the Weyl fermion multiplets of the
model. The four-vectors $A_\mu^a$, where the index $a$ runs over 
$n_{\text{gauge}}$ gauge bosons in the adjoint representation of the
corresponding gauge group, denotes the gauge bosons of the
model. $-\mathcal{L}_S$ denotes the extended Higgs potential
(including the SM Higgs parts). This consists of the terms
$L^i, L^{ij}, L^{ijk}, L^{ijkl}$ and the real scalar fields $\Phi_i$,
with $i, j, k, l = 1, \dots , n_{\text{Higgs}}$. $Y^{IJk}$, with
$I , J = 1 \dots n_{\text{fermion}}$, are the couplings for the
interactions between the scalar and the fermionic fields. $G^{abij}$,
with $a,b = 1\dots n_{\text{gauge}}$, are the couplings for the
interactions between the scalar and the bosonic fields.

After symmetry breaking the scalar fields are expanded around there
VEVs, $\omega_i$ as
\begin{equation}
\Phi_i(x) =  \omega_i + \phi_i(x). \label{eq:phiexpand}
\end{equation}
Putting Eq.~\pref{eq:phiexpand} in Eqs.~\pref{Eq:LS}-\pref{Eq:LG} gives
\begin{align}
-\mathcal{L}_S &= \Lambda + \Lambda^i_{(S)} \phi_i + \frac{1}{2}
-\Lambda_{(S)}^{ij} \phi_i\phi_j + \frac{1}{3!} \Lambda^{ijk}_{(S)}
-\phi_i\phi_j\phi_k + \frac{1}{4!}
-\Lambda_{(S)}^{ijkl}\phi_i\phi_j\phi_k\phi_l \\ 
-\mathcal{L}_F &= \frac{1}{2} M^{IJ} \Psi_I\Psi_J + \frac{1}{2}
-Y^{IJk}\Psi_I\Psi_J\phi_k + c.c.  \\ 
\mathcal{L}_G &= \frac{1}{2} \Lambda^{ab}_{(G)} A_{a\mu}A_b^{\mu}
-+\frac{1}{2} \Lambda^{abi}_{(G)} A_{a\mu}A_b^{\mu}\phi_i +
-\frac{1}{4} \Lambda^{abij}_{(G)} A_{a\mu}A_{b}^\mu\phi_i\phi_j \;,
\end{align}
where 
\begin{align}
\Lambda &= V^{(0)}(\omega_i) = L^i \omega_i + \frac{1}{2!}
L^{ij}\omega_i \omega_j + \frac{1}{3!} L^{ijk} \omega_i \omega_j
\omega_k + \frac{1}{4!} L^{ijkl} \omega_i  
\omega_j \omega_k \omega_l \label{Vtreelevel} \\
\Lambda_{(S)}^i &= L^i + L^{ij} \omega_j + \frac{1}{2} L^{ijk}\omega_j
\omega_k  + \frac{1}{6} L^{ijkl}\omega_j\omega_k\omega_l \, , \quad \Lambda_{(S)}^{ij} = L^{ij} + L^{ijk}\omega_k + \frac{1}{2} L^{ijkl}\omega_k\omega_l  \label{eq:scalarten},  \\ 
\Lambda_{(S)}^{ijk} &= L^{ijk}+L^{ijkl}\omega_l  \, , \quad \Lambda_{(S)}^{ijkl} = L^{ijkl}
\, , \quad \Lambda_{(G)}^{ab} = \frac{1}{2} G^{abij}\omega_i\omega_j \label{Gabterm}
\, , \quad \Lambda_{(G)}^{abi} = G^{abij}\omega_j\\
\Lambda_{(G)}^{abij} &= G^{abij} \label{Gabijterm} \\
\Lambda_{(F)}^{IJ} &= M^{\ast IL} M_{L}^{\; J} = Y^{\ast  ILk}Y_L^{\; Jm} \omega_k\omega_m \;, \quad \mbox{with} M^{IJ} = Y^{IJk}\omega_k. \label{Fterm} \\
\end{align}

We compute each of the terms $\omega_i, \, L^i, \, L^{ij}, \, L^{ijk},
\, L^{ijkl}, \, G^{abij}$ and $Y^{IJk}$ in {\tt C++} format and then
develop the IDM model files needed for BSMPT to work. 

\paragraph{The Coleman-Weinberg part of the effective potential}
Radiative quantum corrections affects the vacuum structure of
potentials at the loop levels. This is accounted for using the 1-loop
correction known as Coleman-Weinberg potential~\cite{Coleman:1973jx}
given by 
\begin{equation}
V_{CW}(\omega) = \frac{1}{4 \, \left(4\pi\right)^2}  \sum_{X={S,G,F}}(-1)^{2s_X} (1+2s_X) Tr[(\Lambda^{xy}_{(X)})^2 ( \log(\frac{1}{\mu^2} \Lambda^{xy}_{(X)} ) - k_X)] \label{CWpotential},
\end{equation}
where $s_X$ represents the spin of the field $X$. $X=S,G$ and $F$,
respectively, represent scalar, gauge and fermionic fields. The
indices $xy$ correspond to the scalar indices $ij$, the gauge indices 
$ab$ and the fermion indices $IJ$ for $X=S,G$ and $F$, respectively.
The sum over $X$ is for all degrees of freedom including colour for
the quarks. $\Lambda_{(S)}^{ij}$, $\Lambda_{(G)}^{ab}$ and
$\Lambda_{(F)}^{IJ}$ are as given in
Eqs.~\pref{eq:scalarten},~\pref{Gabterm} and ~\pref{Fterm}. The
$\overline{\mbox{MS}}$ renormalisation scheme constants are
\beq
k_X = \left\{ \begin{array}{ll} \frac{5}{6} \;, & \quad \mbox{for
                                                  gauge bosons}
\\[0.1cm] \frac{3}{2} \;, & \quad \mbox{otherwise}
\end{array} \right. 
\eeq
The renormalisation scale $\mu$ is set to the SM Higgs multiplet VEV
at $T=0$, $\mu = v(T=0) \approx 246.22 \textrm{ GeV}$.  

\paragraph{The counter term part of the effective potential}
The BSMPT package was designed to use loop-corrected masses and
mixing angles as input. As such, the $\overline{\mbox{MS}}$
renormalisation scheme used for the Coleman-Weinberg part of the
effective potential has to the modified into the on-shell
renormalisation scheme. It is for this reason that the counter term
part of the effective potential, $V_{\text{CT}}$, is
added. $V_{\text{CT}}$ is obtained by replacing bare parameters
$p^{(0)}$ of the tree-level potential $V^{(0)}$ by the renormalised
ones, $p$, and their corresponding counter terms $\delta p$
\begin{align}
V^{\text{CT}} &= \sum_{i=1}^{n_p} \frac{\partial V^{(0)}}{\partial
  p_i} \delta p_i +  \sum_{k=1}^{n_v} \delta T_k \left(\phi_k +
\omega_k \right) \label{vcounterterm}. 
\end{align}
Here $n_p$ is the number of parameters of the potential.
$\delta T_k$ represent the counter terms of the tadpoles
$T_k$corresponding to the $n_v$ directions in field space
with non-zero VEV.

\paragraph{The thermal corrections}
The temperature dependent part of the effective potential $V^{(T)}$ is
given by 
\cite{Dolan:1973qd,Quiros:1999jp} 
\begin{align}
V^T (\omega,T) &= \sum{X={S,G,F}}{} (-1)^{2 s_X}
(1 + 2 s_X)\frac{T^4}{2\pi^2} J_{\pm}\left(\Lambda^{xy}_{(X)}/T^2
\right)\,, 
\end{align}
where $J_{\pm}$ is for bosons or fermions respectively, 
\begin{align}
J_{\pm}\left(\Lambda_{(X)}^{xy}/T^2\right) &= \textrm{Tr}\left[ \int_{0}^{\infty} \,dk\,  
k^2 \log\left[ 1 \pm \exp\left( -\sqrt{k^2 + \Lambda^{xy}_{(X)}/T^2}\right) \right] \right]
\,. 
\end{align}
Taking the finite temperature effect, the daisy
corrections~\cite{Carrington:1991hz} to the scalar and gauge boson
masses are also implemented in the BSMPT package for the IDM model. 

\section{Implementation of the IDM model to BSMPT} 
\label{makeBSMPT}
New models can be implemented in BSMPT. For the IDM, the Lagrangian density terms are written in the required format as described in Appendix~\ref{veffective}. Using 
$\Phi_i = \lbrace h , x_1 , x_2 , x_3 , y_1 , y_2 , S , R \rbrace$, the code needs 
$\left\lbrace L^i,L^{ij},L^{ijk},L^{ijkl},Y^{IJk},G^{abij}\right\rbrace$ as in
Eqs.~\pref{Eq:LS}, \pref{Eq:LF}, and \pref{Eq:LG} specified in {\tt C++} form.
For instance $L^{ij} = 0$ unless for $i=j$ for which we have: 
\begin{align}
L^{i,j=x_{3}}=L^{i,j=x_{2}}=L^{i,j=x_{1}}=L^{i,j=h}= -\mu^2_{h} \\
L^{i,j=R}=L^{i,j=S}= L^{i,j=y_{2}}=L^{i,j=y_{1}}= \mu^2_{Q}.
\end{align}

The same has to be done for the counter terms
  \begin{eqnarray}
    V^{CT} = \delta\mu^2_{h}H^{\dagger}H + \delta\mu^2_{Q}Q^{\dagger}Q +
    \delta\lambda_1 \left(H^{\dagger}H\right)^2 +
     \delta\lambda_2
    [(QQ)_1(\overline{Q}\,\overline{Q})_1]_{0} +
    \delta\alpha(H^{\dagger}H)(Q^{\dagger}Q) + \nonumber\\
    \delta\beta[(\overline{H}H)_{1}(\overline{Q}Q)_{1}]_{0}
    + \Big\{\delta\kappa_1[(HH)_{1}(\overline{Q}\,\overline{Q})_{1}]_{0} +    
    H.c.\Big\} + \delta T \left(h + \textrm{VEV}\right). 
  \end{eqnarray}
In order to add this part of the IDM effective Lagrangian to the BSMPT package, 
the coefficients
$\left\lbrace L^i,L^{ij},L^{ijk},L^{ijkl},Y^{IJk},G^{abij}\right\rbrace $
have to be computed symbolically and then written in {\tt C++} form. 
Applying on-shell renormalisation leads to the equations
\begin{eqnarray}
\left.\partial_{\phi_i} V^{\text{CT}}\right|_{\phi =\langle \phi^c\rangle_{T=0}} &=&
-\left.\partial_{\phi_i} V^{\text{CW}}\right|_{\phi =\langle \phi^c \rangle_{T=0}} \\
\left.\partial_{\phi_i}\partial_{\phi_j} V^{\text{CT}}
\right|_{\phi=\langle \phi^c \rangle_{T=0}} &=& -\left.\partial_{\phi_i}
\partial_{\phi_j} V^{\text{CW}} \right|_{\phi=\langle\phi^c \rangle_{T=0}}.  \
\end{eqnarray}
Solving these equations with respect to the counter terms gives
 \begin{eqnarray}
  \delta\mu^2_{h} &=& -\frac{1}{2} H^{\text{CW}}_{h,h} + \frac{3}{2} H^{\text{CW}}_{x_3,x_3} \\
  \delta\mu^2_{Q} &=& -\frac{1}{2} H^{\text{CW}}_{y_2,y_2} - \frac{1}{4} H^{\text{CW}}_{S,S} - \frac{1}{4} H^{\text{CW}}_{R,R}\\
  \delta\lambda_{1} &=& \frac{1}{2v^2}\left( -H^{\text{CW}}_{h,h} + H^{\text{CW}}_{x_3,x_3}\right) \\
  \delta\lambda_{2} &=& 0\\
  \delta K &=& \frac{-\sqrt{3}}{2v^2}\left( H^{\text{CW}}_{S,S} + H^{\text{CW}}_{R,R}\right) \\
  \delta\alpha &=& 0\\
  \delta\beta &=& \frac{\sqrt{3}}{v^2}\left( H^{\text{CW}}_{S,S} + H^{\text{CW}}_{R,R} -  2H^{\text{CW}}_{y_2,y_2}\right)\\
  \delta T &=& vH^{\text{CW}}_{x_3,x_3} - N^{\text{CW}}_{h}
\end{eqnarray}

 Finally for models with a different Yukawa and gauge sectors relative to the SM ones, the thermal corrections codes of the BSMPT has to be modified. For the IDM, the gauge sector differs. To account for this, we modified the function {\tt CalculateDebyeGaugeSimplified()} using
\begin{equation}
\Pi_{(G)}^{ab} = T^2 \frac{2}{3} \left(\frac{\tilde{n}_H}{8} + 5 \right) \frac{1}{\tilde{n}_H}
\sum\limits{m=1}{n_{\text{Higgs}}} \Lambda^{aamm}_{(G)} 
\delta_{ab}
\end{equation} 
where $\Pi_{(G)}^{ab}$ belongs to daisy correction to thermal masses of gauge bosons, $\tilde{n}_H$ represents the number of Higgs bosons coupled to the SM gauge sector. For the IDM, this leads to 
\begin{align}
\Pi^{a,b=W_{0}}_{G}= \Pi^{a,b=W_{1}}_{G}= \Pi^{a,b=W_{2}}_{G}=2g^2 \\
\Pi^{a,b=B_{0}}_{G}= 2{g^{\prime}}^{2} 
\end{align}
where, $g$ and $g^\prime$ are SM $SU(2)_L$ and $U(1)$ gauge couplings. 
\end{appendix}

\end{document}